\documentclass[aps,superscriptaddress,twocolumn,amsmath,amssymb,titlepage]{revtex4-1}

\usepackage{bbold}
\usepackage{mathptmx}
\usepackage{psfrag,graphicx}
\usepackage{dcolumn}
\usepackage{amsmath,amssymb}
\usepackage{bm}
\usepackage{color}
\usepackage{latexsym}
\usepackage{epstopdf}
\usepackage{color}
\usepackage[english]{babel}
\usepackage{latexsym}
\usepackage{psfrag,graphicx}
\usepackage{subfig}
\usepackage{amsmath}
\usepackage{amssymb}
\usepackage{amsfonts}
\usepackage{bm}
\usepackage{natbib}
\usepackage{epstopdf}
\usepackage{appendix}
\usepackage{rotating}

\definecolor{mygrey}{gray}{0.35}
\definecolor{myblue}{rgb}{0.2,0.2,0.8}
\definecolor{myzard}{cmyk}{0,0,0.05,0}
\definecolor{mywhite}{rgb}{1,1,1}
\definecolor{mywhite}{rgb}{1,1,1}
\definecolor{myred}{rgb}{1,0.,0.3}

\usepackage[colorlinks=true,citecolor=myblue,linkcolor=myred]{hyperref}

\def\ba{\begin{align}}
\def\enda{\end{align}}
\def\bi{\begin{itemize}}
\def\ei{\end{itemize}}

\def\be{\begin{equation}}
\def\ee{\end{equation}}
\def\bea{\begin{eqnarray}}
\def\eea{\end{eqnarray}}
\def\bse{\begin{subequations}}
\def\ese{\end{subequations}}

\newcommand{\ket}[1]{|{#1}\rangle}                       
\newcommand{\average}[1]{\langle {#1} \rangle}           

\newcommand{\Ignore}[1]{ }

\begin{document}

\title{Dzyaloshinskii-Moriya and dipole-dipole interactions affect coupling-based Landau-Majorana-St\"uckelberg-Zener transitions}

\author{R. Grimaudo}
\affiliation{ Dipartimento di Fisica e Chimica dell'Universit\`a di Palermo, Via Archirafi, 36, I-90123 Palermo, Italy}

\author{H. Nakazato}
\affiliation{Department of Physics, Waseda University, Tokyo 169-8555, Japan}
\affiliation{Institute for Advanced Theoretical and Experimental Physics, Waseda University, Tokyo 169-8555, Japan}

\author{A. Messina}
\affiliation{ Dipartimento di Matematica ed Informatica dell'Universit\`a di Palermo, Via Archirafi, 34, I-90123 Palermo, Italy}

\author{ N. V. Vitanov }
\affiliation{Department of Physics, St. Kliment Ohridski University of Sofia, 5 James Bourchier Boulevard, 1164 Sofia, Bulgaria}

\begin{abstract}
It has been theoretically demonstrated that two spins (qubits or qutrits), coupled by exchange interaction only, undergo a coupling-based joint Landau-Majorana-St\"uckelberg-Zener (LMSZ) transition when a linear ramp acts upon one of the two spins. Such a transition, under appropriate conditions on the parameters, drives the two-spin system toward a maximally entangled state. In this paper, effects on the quantum dynamics of the two qudits, stemming from the Dzyaloshinskii-Moriya (DM) and dipole-dipole (d-d) interactions, are investigated qualitatively and quantitatively. The enriched Hamiltonian model of the two spins, shares with the previous microscopic one the same C2-symmetry which once more brings about an exact treatment of the new quantum dynamical problem. This paper transparently  reveals that the DM and d-d interactions generate independent, enhancing or hindering, modifications in the dynamical behaviour predicted for the two spins coupled exclusively by the exchange interaction. It is worthy noticing that, on the basis of the theory here developed, the measurement of the time evolution of the magnetization in a controlled LMSZ scenario, can furnish information on the relative weights of the three kinds of couplings describing the spin system.
This possibility is very important since it allows in principle to legitimate the choice of the microscopic model to be adopted in a given physical scenario. 
 
\end{abstract}

\date{\today}

\pacs{ 
PACS 
}

\maketitle

\section{Introduction}

The anisotropic interaction term known as Dzyaloshinskii-Moriya (DM) interaction or anti-symetric exchange interaction has been firstly phenomenologically introduced by I. Dzyaloshinskii to understand and explain the weak ferromagnetism in antiferromagnetic crystals \cite{Dzyaloshinskii}.
Later, T. Moriya has furnished a theoretical derivation of such an interaction term grounded on a robust general theory of the superexchange interaction including the spin-orbit coupling \cite{Moriya}.
This anisotropic interaction term arises when antiferromagnetic systems present low symmetry and strongly depends on the geometry of the system as shown by T. Moriya \cite{Moriya}.
Mathematically it is written precisely as
\begin{equation}
\mathbf{d} \cdot (\mathbf{S}_1 \times \mathbf{S}_2)
\end{equation}
where $\mathbf{S}_1$ and $\mathbf{S}_2$ are the two interacting spins and $\mathbf{d}$ is the so called DM vector whose orientation is determined by the geometry and then by the symmetry of the spin system.

Generally, the DM interaction is small with respect to other types of interactions, such as the exchange or the dipole-dipole interaction.
In spite of this, however, its contribution proved to be essential to correctly describe several antiferromagnetic molecules \cite{Dender,Kohgi,Tsukada}.
Moreover, the DM interaction proves to be crucial in figuring out the correct dynamics of complex spin systems, e.g. tunneling and interference between energy levels in single-magnet molecules \cite{Caretta,Ramsey}.

It is interesting to point out that the DM interaction plays an important role for quantum computing applications \cite{Wu} and in several different contexts like quantum dots \cite{Kavokin}, and spin chain dynamics \cite{Jafari,Wang}.
It has been demonstrated that the presence of the DM interaction in spin chains deeply influences several different physical quantities, like Berry's phase \cite{Kwan}, quantum phase interference \cite{Wernsdorfer}, quantum phase transitions \cite{LiuKong}, entanglement transfer \cite{Maruyama}, thermal entanglement and teleportation \cite{Zhang} and classical and quantum correlations \cite{Liu}.
The effects of the DM interaction on the entanglement formation in spin-qubit chains is particularly relevant.
This interaction, in fact, fosters the occurrence of long-distance correlations \cite{Sahling,Jurcevic,Richerme,Boness} which turn out to be crucial in quantum technological applications.
Since the entanglement is a key resource in quantum information \cite{Amico}, its controllable production and manipulation \cite{Arnesen} is, indeed, of fundamental importance.

The scope of this paper is to analyse the effects of both the DM interaction and the dipole-dipole (d-d) interaction in the dynamics of a two-qubit system.
The d-d interaction between two spin sites takes into account the overlap between the wave functions of the two interacting systems and is usually written in terms of spin variables as \cite{Bolton}
\begin{equation}
\mathbf{S}_1 \cdot \mathbf{D} \cdot \mathbf{S}_2,
\end{equation}
where $\mathbf{D}$ is the traceless dipole tensor.
The form of this tensor, that is, its non-vanishing entries strictly depend on the symmetries of the system.
Systems with uniaxial symmetry, for example, would be characterized by a diagonal tensor with $d_{xx}=d_{yy}=-d_{zz}/2$ \cite{Bolton}, where $d_{kk} ~ (k=x,y,z)$ stands for the diagonal term in the $k$ direction.

Differently from the papers cited above, we take into account the presence of an external time-dependent field.
In a previous work \cite{GVMqubits} the authors have studied the same problem considering only the exchange interaction.
In this typical scenario, they have demonstrated the possibility of generating maximally entangled states of the two spins by applying a Landau-Majorana-St\"uckelberg-Zener (LMSZ) ramp \cite{LMSZ}.
In this paper, we want to investigate how the DM and the d-d interactions influence such a dynamic behaviour and to highlight intriguing effects which can arise from the interplay of the two types of interactions and the external fields.

The choice of the LMSZ framework \cite{LMSZ} is based on the fact that it is a very well known and quite easily experimentally realizable time-dependent scenario.
Despite to be idealized (infinite time duration of the experimental procedure implying divergent energies of the adiabatic states), the LMSZ model succeeds in grasping relevant physical features of the physical systems under scrutiny \cite{Shevchenko1}.
However, if one is interested in a more realistic situation, it is worth underlining that the exact solution of the dynamical problem is known also for a finite time windowed LMSZ procedure \cite{Vit-Garr}.
This aspect, then, allows us to make predictions closer, at least in principle, to the experimental results.
In this respect, it is important to stress that realistic descriptions of quantum systems subjected to the LMSZ scenario are required to include also environmental effects.
To this end several analysis and models have been proposed \cite{Akulin,Vitanov,Ivanov,Pok-Sin2003,Militello2,Militello3}.
Different sources of incoherences in the experimental framework, indeed, can influence the system dynamics \cite{Petta,Anderlini,Foletti,Das Sarma Nat}, like relaxation processes (e.g., spontaneous emission) or interaction with a surrounding environment (e.g., nuclear spin bath).
Also such an aspect has been taken into account in the previous paper \cite{GVMqubits} and is analysed here too in presence of the DM interaction.

In this work we study the effects of the DM and d-d interactions in a system of two interacting three-level systems too, where the occurrence of coupling-based LMSZ transitions has been analytically demonstrated \cite{GVMqutrits}. 
In the last reference, in fact, the same authors demonstrated the occurrence of coupling-based LMSZ transitions in case of two interacting spin-qutrits.
The interest toward systems of interacting $N$-level systems has grown exponentially in the last years most of all thanks to the wide range of applications in quantum information.
Indeed, methods to manipulate qutrits \cite{Klimov2003,Vitanov2012} and qudits \cite{Ivanov2006} have been developed.
Moreover, qutrit systems have proven to be very useful in developing a more secure quantum communication \cite{Cerf2002}, besides showing a huge potential in generating new type of entanglement \cite{Vaziri2002}.

The paper is organised as follows.
Section \ref{Qubits} deals with two interacting spin-qubit systems.
In this section, the generalization of the qubit model studied in Ref. \cite{GVMqubits}, including the DM and d-d interaction terms, is presented.
Detectable dynamical effects stemming from the presence of both or just one of the interactions taken into account are brought to light.
Furthermore, DM and d-d-induced changes on the possibility to get entangled states of the two qubits through the coupling-based LMSZ transitions are discussed.
An analogous analysis is developed in Sec. \ref{Qutrits} for a two-spin-qutrit system.
Conclusive Remarks are reported in the last section \ref{Sec C}.

\section{Spin-Qubits}\label{Qubits}

\subsection{The model and the coupling-based LMSZ effect enhancement}

Let us consider the following two-qubit model
\begin{equation} \label{H DM}
\begin{aligned}
{H} =&
\hbar\omega_{1}(t)\hat{\sigma}_{1}^{z}+\hbar\omega_{2}(t)\hat{\sigma}_{2}^{z}+\gamma_{x}\hat{\sigma}_{1}^{x}\hat{\sigma}_{2}^{x}+\gamma_{y}\hat{\sigma}_{1}^{y}\hat{\sigma}_{2}^{y}+\gamma_{z}\hat{\sigma}_{1}^{z}\hat{\sigma}_{2}^{z} \\
&+\gamma_{xy}\hat{\sigma}_1^x\hat{\sigma}_2^y+\gamma_{yx}\hat{\sigma}_1^y\hat{\sigma}_2^x.
\end{aligned}
\end{equation}
The two spin-1/2's are subjected to two local time-dependent magnetic fields along the $z$ axis and the first three coupling terms account for an anisotropic exchange interaction \cite{Bolton}.
The last two coupling terms, instead, represent the contribution of both the anisotropic d-d interaction and DM interaction \cite{Bolton}.
We point out that the presence of d-d interaction implies that we have an equal contribution of the two mixed terms ($\hat{\sigma}_1^x\hat{\sigma}_2^y$ and $\hat{\sigma}_1^y\hat{\sigma}_2^x$), that is, $\gamma_{xy}=\gamma_{yx}$ \cite{Bolton}.
The DM interaction, instead, arises when we have opposite contributions in the two mixed terms, namely $\gamma_{xy}=-\gamma_{yx}$  \cite{Bolton}.
Therefore, considering both the two interactions, we can write two independent parameters ($\gamma_{xy}$ and $\gamma_{yx}$) for the two coupling terms under consideration.

The interaction model here considered is appropriate for a system possessing a $C_2$-symmetry with respect to the $z$ axis \cite{GMN}.
It is possible to convince oneself easily, indeed, that the Hamiltonian keeps its form after a rotation of $\pi$ around the $z$ axis for both spins (namely it means to perform the following transformation $\hat{\sigma}_k^x \rightarrow -\hat{\sigma}_a^x$, $\hat{\sigma}_a^y \rightarrow -\hat{\sigma}_a^y$, $\hat{\sigma}_a^z \rightarrow \hat{\sigma}_a^z$, $a=1,2$) \cite{GMN}.

Thanks to such a $C_2$-symmetry it is easy to identify the constant of motion $\hat{\sigma}_{1}^{z}\hat{\sigma}_{2}^{z}$ which implies the existence of two dynamically invariant two-dimensional subspaces (related to the two eigenvalues $\pm$ of the integral of motion) \cite{GMN}.
In this way the dynamical problem of the two spin-1/2's may be traced back to the solution of two independent problems of single (fictitious) spin-1/2 governed by the two Hamiltonians \cite{GMN}
\begin{equation}
\begin{aligned}
H_{\pm}=
\hbar\Omega_\pm(t)\hat{\sigma}^z+\gamma_\pm\hat{\sigma}^x+\Gamma_\pm\hat{\sigma}^y\pm\gamma_z\mathbb{1}_{\pm},
\end{aligned}
\end{equation}
with $\Omega_\pm(t)=\omega_1\pm\omega_2$, $\gamma_\pm=\gamma_x\mp\gamma_y$, $\Gamma_\pm=\pm\gamma_{xy}+\gamma_{yx}$ and $\mathbb{1}_\pm$ being the two-dimensional identity operator related to each subspace.

Let us consider now an LMSZ-like scenario, as done in Ref. \cite{GVMqubits},
\begin{equation}
\hbar\omega_1(t)=\alpha~t, \quad \hbar\omega_2(t)=0,
\end{equation}
that is, an LMSZ ramp (a linearly varying magnetic field) is applied only on the first spin.
If we study the two-qubit LMSZ transitions in each subspaces, it is easy to convince oneself that they result in
\begin{equation}\label{P+- gen}
P_\pm=|\average{\pm\pm|U_+(\infty)|\mp\mp}|^2=1-\exp\{ -2\pi(\gamma_\pm^2+\Gamma_\pm^2)/\hbar\alpha \},
\end{equation}
where the single-qubit states are defined as $\sigma^z\ket{\pm}=\pm\ket{\pm}$.
The analogous expressions in the absence of the d-d and DM interactions, obtained in Ref. \cite{GVMqubits}, are reported in Eqs. \eqref{P+ P- simple}.
It is worthy noticing that this result shows that the effect of the presence of new Hamiltonian terms due to the DM and d-d interactions is to increase the probability of the two-qubit LMSZ transition.
Thus, the physical effect of coupling-assisted LMSZ transitions for the two-qubit system, brought to light in Ref. \cite{GVMqubits}, turns out to be strengthened by the presence of the new interaction terms.
We point out that the physical reason of the appearance of the term $\sqrt{\gamma_\pm^2+\Gamma_\pm^2}$ relies on the fact that this term is exactly the modulus of the real transverse magnetic field we get by rotating the Hamiltonian $H_\pm$ around the $z$-axis by the angle $\arctan(-\Gamma_\pm/\gamma_\pm)$.

\subsection{Effects of the DM and d-d couplings on the dynamics}

It is important to notice that the relative weights of the different types of interaction make the Hamiltonian possess different symmetries, giving rise to different dynamics and consequently to different physical effects.
In the following we take into account all possible physical scenarios related to different specific interactions.

If we had an isotropic exchange interaction, that is, $\gamma_x=\gamma_y=\gamma/2$, we would obtain
\begin{equation}\label{P iso change + dd and DM}
\begin{aligned}
P_+&=1-\exp\{ -2\pi\Gamma_+^2/\hbar\alpha \}, \\
P_-&=1-\exp\{ -2\pi(\gamma^2+\Gamma_-^2)/\hbar\alpha \}.
\end{aligned}
\end{equation}
In this case, thus, we see that the contribution of DM and d-d interactions is fundamental for the occurrence of LMSZ transition from $\ket{--}$ to $\ket{++}$.
We would get the analogous result if we were in presence only of d-d interaction without the DM contribution.
In this instance, indeed, we would have $\gamma_{xy}=\gamma_{yx}=\Gamma/2$ leading to
\begin{equation}
\begin{aligned}\label{P iso change + dd}
P_+&=1-\exp\{ -2\pi\Gamma^2/\hbar\alpha \}, \\
P_-&=1-\exp\{ -2\pi\gamma^2/\hbar\alpha \}.
\end{aligned}
\end{equation}
Further, if the system were characterized by just a pure DM interaction without a relevant contribution of the anisotropic d-d interaction, meaning that $\gamma_{xy}=-\gamma_{yx}=\Gamma/2$, we would get
\begin{equation}\label{P iso change + DM}
\begin{aligned}
P_+=0, \qquad P_-=1-\exp\{ -2\pi(\gamma^2+\Gamma^2)/\hbar\alpha \}.
\end{aligned}
\end{equation}
Thus, with an isotropic exchange interaction, the presence of only DM coupling does not generate an LMSZ transition in the first subdynamics involving $\ket{++}$ and $\ket{--}$.

Finally, considering an anisotropic exchange, we get
\begin{equation}\label{P aniso change + dd}
\begin{aligned}
P_+&=1-\exp\{ -2\pi(\gamma_+^2+\Gamma^2)/\hbar\alpha \}, \\
P_-&=1-\exp\{ -2\pi\gamma_-^2/\hbar\alpha \}.
\end{aligned}
\end{equation}
in case of d-d interaction only and we have
\begin{equation}\label{P aniso change + DM}
\begin{aligned}
P_+&=1-\exp\{ -2\pi\gamma_+^2/\hbar\alpha \}, \\
P_-&=1-\exp\{ -2\pi(\gamma_-^2+\Gamma^2)/\hbar\alpha \},
\end{aligned}
\end{equation}
when only the DM contribution is considered.

These examples illustrate how essential are both the symmetries and the anisotropies in determining the dynamics  and the response of the physical system when the system is subjected to externally applied fields.
We emphasize, moreover, that the physical effects highlighted above are relatively easily observable in laboratory during an LMSZ experiment.
From Eqs. \eqref{P iso change + dd and DM}, \eqref{P iso change + dd} and \eqref{P iso change + DM}, we see, in fact, that, by analysing the transition $\ket{--} ~ \rightarrow ~ \ket{++}$, we may get the transition or not depending on the relative weights of the interaction terms.
Therefore, by measuring the time-behaviour of a physical observable like the magnetization of the system we can have a direct confirmation of the occurrence or non-occurrence of the transition and then we can understand what kind of interaction must be taken into account  in the Hamiltonian model describing the physical system.
The characteristic time-scale of the transition, instead, would reveal the magnitude of the LMSZ parameter ruling the dynamics (that is of the exponent of the LMSZ transition expression) giving us information about the weights of the interactions.
We point out that for an ideal LMSZ scenario it is not properly correct to talk about time-scale because of the infinite time-duration of the procedure.
However, what we refer to by saying time-scale is a more concrete and physically reasonable scenario consisting in a time-windowed procedure where time-scale effects arise indeed and can be brought to light.

\subsection{Entanglement}

A remarkable aspect related to the coupling-based two-qubit LMSZ transitions concerns the possibility of generating maximally entangled states of the two spins.
As shown in Ref. \cite{GVMqubits}, when the system is initialized in $\ket{--}$ or $\ket{-+}$, the two asymptotic concurrence \cite{Wootters} curves, respectively writeable as
\begin{subequations}
\begin{align}
C_+&=2|c_{++}c_{--}|=2\sqrt{P_+(1-P_+)},\label{C1a} \\
C_-&=2|c_{+-}c_{-+}|=2\sqrt{P_-(1-P_-)},\label{C1b}
\end{align}
\end{subequations}
($c_{++}$ and $c_{--}$ ($c_{+-}$ and $c_{-+}$) are the asymptotic amplitudes of the states $\ket{++}$ and $\ket{--}$ ($\ket{+-}$ and $\ket{-+}$), respectively) reach the maximum value $C_\pm=1$ when $P_\pm=1/2$, respectively.
This circumstance happens when the exponential term in Eqs. \eqref{P+- gen} assumes the value $1/2$, implying then
\begin{equation}\label{Condition max ent}
2\pi(\gamma_\pm^2+\Gamma_\pm^2)/\hbar\alpha=\ln(2).
\end{equation}

This result means that, when the condition in Eq. \eqref{Condition max ent} is fulfilled, the two-qubit system asymptotically reaches the maximally entangled state $(\ket{++}+e^{i\phi}\ket{--})/\sqrt{2}$ in the first case and $(\ket{+-}+e^{i\phi'}\ket{-+})/\sqrt{2}$ in the second case.
This fact was proved in Ref. \cite{GVMqubits} by studying the exact time-dependence of the concurrence exploiting the analytical solutions of the time-windowed LMSZ dynamical problem reported in Ref. \cite{Vit-Garr}.
In that case only the exchange interaction was taken into account; in the case considered in this work the analysis is analogous with the only difference, however, that the LMSZ parameter ruling the dynamics includes the contribution stemming from the presence of the DM and the d-d interactions.

It is worthy underlining that the maximally entangled state generation requires \emph{non-adiabatic} conditions.
This circumstance is due to the fact that the entanglement maximization in each subspace requires $P_\pm=1/2$, that is, a half transition, in order to get an equally weighted superposition of the two involved states.
Conversely, adiabatic conditions, in fact, ensure the full transition ($P_\pm=1$) from $\ket{--}$ ($\ket{-+}$) to $\ket{++}$ ($\ket{+-}$).
It is important to point out that this circumstance, that is, the non-adiabatic entanglement generation is strictly related to the fact that we are considering a full level crossing.
If we consider the `half crossing', i.e., the case in which the evolution starts exactly at the crossing, then the LMSZ transition probability $1-\exp(-\Lambda)$ is replaced by $[1-\exp(-\Lambda/2)]/2$ \cite{Vit-Garr,Carr-Hioe}.
We see, then, that the half-crossing dynamics allows us to obtain the same results, namely a half transition and the consequent entanglement generation for the two-spin system, maintaining \emph{adiabatic} conditions.
This last aspect, which opens possible interesting applications for two-qubit scenarios, strengthens the physical relevance of the half-crossing dynamics, already witnessed by deep theoretical insights \cite{Vitpra99,Vit-Zlat} and successful uses in experiments \cite{Yamazaki,Randall}.

Finally, we point out that, differently from what happens for the full LMSZ transitions, the presence of DM and d-d interactions could promote or hinder the appearance of entanglement.
In fact, as far as the full transitions are concerned, if the adiabatic conditions are satisfied by the exchange parameters, namely $\gamma_\pm^2/\hbar\alpha \gg 1$, then the two interactions increase the transition probability or, in other words, make the characteristic time-scale of the transition shorter.
On the other hand, if $\gamma_\pm^2/\hbar\alpha \gg 1$ are not satisfied, DM and d-d interaction terms have the effect to enhance the exponential ratio and then to foster adiabatic conditions and a consequent full transition.
In this last case, thus, the presence of the two types of interactions has only positive effects.

As far as the entanglement is concerned, instead, if in a given situation, we have $2\pi\gamma_\pm^2/\hbar\alpha \simeq \ln(2)$, the DM and/or the d-d interactions negatively influence its occurrence since the numerical value of the ratio in Eq. \eqref{Condition max ent} would be different from the required one for  the half transition.
Interestingly, it could rather happen that if $2\pi\gamma_\pm^2/\hbar\alpha < \ln(2)$ the two interactions positively contribute to reach the necessary condition for the maximally entangled state generation.
Considering the entanglement, therefore, both constructive and destructive physical effects can stem from the presence of DM and d-d interactions.

\section{Spin-Qutrits} \label{Qutrits}

\subsection{The model and the coupling-based two-qutrit LMSZ transition enhancement}

We consider now the same model analysed before for two interacting three-level systems \cite{GVMqutrits}, namely
\begin{eqnarray} 
H=
\hbar\omega_{1}\hat{\Sigma}_{1}^{z}
+\gamma_{x}\hat{\Sigma}_{1}^{x}\hat{\Sigma}_{2}^{x}
+\gamma_{y}\hat{\Sigma}_{1}^{y}\hat{\Sigma}_{2}^{y}
+\gamma_{xy}\hat{\Sigma}_{1}^{x}\hat{\Sigma}_{2}^{y}+\gamma_{yx}\hat{\Sigma}_{1}^{y}\hat{\Sigma}_{2}^{x}.
\end{eqnarray}
The Hamiltonian fulfills the condition $[H,\hat{K}]=0$ with $\hat{K}=\cos[\pi(\hat{\Sigma}_1^z+\hat{\Sigma}_2^z)]$ \cite{GMIV,GVMqutrits}.
This constant of motion generates two dynamically invariant subspaces: a four-dimensional one and a five-dimensional one \cite{GMIV,GVMqutrits}.

It is worthy noticing that, similarly to what happens when DM and d-d interactions are absent, the four-dimensional two-qutrit subdynamics can be effectively described in terms of two decoupled spin-1/2's whose Hamiltonians read \cite{GMIV}
\begin{subequations}
\begin{align}
&&H_{1}=\frac{\hbar\omega_{1}}{2}\hat{\sigma}_{1}^{z}+
(\gamma_{x}-\gamma_{y})\hat{\sigma}_{1}^{x}+
(\gamma_{xy}+\gamma_{yx})\hat{\sigma}_{1}^{y}, \label{H1}\\
&&H_{2}=\frac{\hbar\omega_{1}}{2}\hat{\sigma}_{2}^{z}+
(\gamma_{x}+\gamma_{y})\hat{\sigma}_{2}^{x}-
(\gamma_{xy}-\gamma_{yx})\hat{\sigma}_{2}^{y}. \label{H2}
\end{align}
\end{subequations}
We see that, compared to the analogous expressions in Eq. \eqref{H1 and H2}, the presence of non diagonal coupling terms due to DM and d-d interactions makes richer the two effective single qubit Hamiltonians by giving rise to the term proportional to $\hat{\sigma}_y$.
What is remarkable here is that analogously to the case analysed in Ref. \cite{GVMqutrits}, also in this instance we can get exact analytical results thanks to the effective description in terms of two decoupled spin-1/2's.
We point out that such a mathematical trick is based on the following mapping \cite{GMIV}
\begin{equation}\label{Mapping}
\begin{aligned}
\ket{10} & \hspace{0,25cm} \leftrightarrow \hspace{0,25cm} \ket{++},\\
\ket{01} & \hspace{0,25cm} \leftrightarrow \hspace{0,25cm} \ket{+-},\\
\ket{0-1} & \hspace{0,25cm} \leftrightarrow \hspace{0,25cm} \ket{-+},\\
\ket{-10} & \hspace{0,25cm} \leftrightarrow \hspace{0,25cm} \ket{--},
\end{aligned}
\end{equation}
between coupled states of qutrits and coupled states of qubits.
The single-qutrit states $\ket{1}$, $\ket{0}$ and $\ket{-1}$ are eigenstates of $\hat{\Sigma^z}$ with eigenvalues $1$, $0$ and $-1$, respectively. 

If two qutrits start from $\ket{-10}$, under an LSMZ ramp $\omega_1=\alpha~t$, they will reach asymptotically the state $\ket{10}$, $\ket{01}$ and $\ket{0-1}$ with probabilities
\begin{equation}\label{LMSZ trans prob 4x4}
\tilde{P_1} \tilde{P_2}, \quad \tilde{P_1}(1-\tilde{P_2}), \quad (1-\tilde{P_1})\tilde{P_2},
\end{equation}
respectively, being
\begin{subequations}\label{Prob fict spin 1/2 qutrits}
\begin{align}
\tilde{P_1}=1-\exp\{-2\pi{(\tilde{\gamma}_-^2+\tilde{\Gamma}_+^2)/\hbar\alpha}\}, \\
\tilde{P_2}=1-\exp\{-2\pi{(\tilde{\gamma}_+^2+\tilde{\Gamma}_-^2)/\hbar\alpha}\}
\end{align}
\end{subequations}
the two probabilities of the two fictitious spin-1/2's accomplishing the down-up transition and $\tilde{\gamma}_\pm=\gamma_{x}\pm\gamma_{y}$, $\tilde{\Gamma}_\pm=\gamma_{xy}\pm\gamma_{yx}$.

From this result we can appreciate that, also in the case of two interacting qutrits, the presence of the DM and d-d interactions increases the two-qutrit LMSZ transitions and consequently reduces the characteristic LMSZ dynamical time scale.
We underline that, similarly to the case of two qubits, the term $\sqrt{\tilde{\gamma}_\pm^2+\tilde{\Gamma}_\pm^2}$ characterizing the LMSZ parameter is the modulus of the actual effective transverse field we get by rotating the Hamiltonian $H_\pm$ around the $z$-axis of the angle $\arctan(-\tilde{\Gamma}_\pm/\tilde{\gamma}_\pm)$.

We take into account now the five-dimensional subdynamics.
As in Ref. \cite{GMIV} it is not possible to deal analytically with the five-dimensional subspace unless specific conditions on the coupling parameters are introduced.
In particular, putting $\gamma_x=\gamma_y=\tilde{\gamma}/2$ (isotropic exchange interaction), $\gamma_{xy}=-\gamma_{yx}=\tilde{\Gamma}/2$ (only DM interaction) and $\gamma_z=0$, we get a further decomposition of the Hilbert subspace in three dynamically invariant subspaces: two unidimensional subspaces and a three-dimensional su(2) one spanned by the two-qutrit standard basis states $\{ \ket{1-1}, \ket{00}, \ket{-11} \}$.
In the latter case, thus, we can describe the two-qutrit dynamics in terms of an effective single three-level system whose Hamiltonian reads
\begin{equation}
H_3={\hbar\omega_1}\hat{\Sigma}_{2}^{z}+\tilde{\gamma}~\hat{\Sigma}_{2}^{x}-\tilde{\Gamma}~\hat{\Sigma}_{2}^{y}.
\end{equation}
When the two spin-qutrits are initialized in the  state $\ket{-11}$, the LMSZ transition probabilities are characterized by the transition probabilities
\begin{equation}\label{LMSZ Tr Pr 3-lev}
\begin{aligned}
P_{-1}^{+1}=\tilde{P}_3^2, \quad
P_{-1}^{0}=2\tilde{P}_3(1-\tilde{P}_3), \quad
P_{-1}^{-1}=(1-\tilde{P}_3)^2,
\end{aligned}
\end{equation}
with $\tilde{P}_3=1-\exp\{-2\pi(\tilde{\gamma}^2+\tilde{\Gamma}^2)/\hbar\alpha\}$, where we labelled with $1,0,-1$ the three states $\{ \ket{1-1}, \ket{00}, \ket{-11} \}$, respectively.
For the fictitious three-level system, we have a generalization of the LMSZ probability expression given by the presence of DM and/or d-d interactions [see Eqs. \eqref{LMSZ Tr Pr 3-lev}].

We underline that we are able to get exact results in the five-dimensional subspace when only the DM interaction is present and when the exchange interaction is isotropic.
This circumstance, of course, generates some changes in the four-dimensional subdynamics which are discussed in the following subsection.
Finally, we point out that the dynamics in the two one-dimensional subspaces is trivial.

\subsection{Dynamical effects of DM and d-d interactions}

Analogously to the qubit case, we can bring to light the effects on the dynamics stemming from the presence of DM and d-d interaction, by considering different physical scenarios.
If only the isotropic exchange interaction were present, that is $\gamma_x=\gamma_y=\tilde{\gamma}/2$ and $\gamma_{xy}=\gamma_{yx}=0$ we would get $\tilde{P}_1=0$ and $\tilde{P}_2=1-\exp\{-2\pi\tilde{\gamma}^2/\hbar\alpha\}$.
We would have an analogous result if the DM interaction is present too, that is $\gamma_{xy}=\gamma_{yx}=\tilde{\Gamma}/2$.
In this case, the only difference appears in the expression of $\tilde{P}_2$ which turns out to be
\begin{equation}
\tilde{P}_2=1-\exp\{-2\pi{(\tilde{\gamma}^2+\tilde{\Gamma}^2)/\hbar\alpha}\}.
\end{equation}
Looking at expressions in Eq. \eqref{LMSZ trans prob 4x4}, we see that, in this instance, the two qutrits have no possibility to pass from the state $\ket{-10}$ to the states $\ket{10}$ and $\ket{01}$.
This circumstance depends on the fact that $\tilde{P}_1=0$ which can be understood considering the fact that $H_1$ cannot generate a transition of the fictitious first qubit since $H_1 \propto \hat{\sigma}^z$, as it is easy to verify by Eq. \eqref{H1}.
The physical reason of this occurrence can be traced back to the further decomposition of the four-dimensional space in two two-dimensional subspaces.
It happens since the conditions on the parameters under scrutiny generate further symmetries on the Hamiltonian operator.
Precisely, the Hamiltonian commutes now with the operator $\hat{S}_T=\hat{S}_1^z+\hat{S}_2^z$ implying a subspace spanned by $\ket{10}$ and $\ket{01}$ and the other one by $\ket{-10}$ and $\ket{0-1}$.
Therefore, if the two-qutrit system is initially prepared in $\ket{-10}$, it can make a transition only towards the state $\ket{0-1}$.

If, instead, besides the isotropic exchange interaction, only the d-d interaction is present ($\gamma_{xy}=\gamma_{yx}=\tilde{\Gamma}/2$), the further symmetry related to the commutator $[H,\hat{S}_T]=0$ is lost.
In this case we get indeed
\begin{subequations}
\begin{align}
&\tilde{P}_1=1-\exp\{-2\pi{\tilde{\Gamma}^2/\hbar\alpha}\}, \\
&\tilde{P}_2=1-\exp\{-2\pi{\tilde{\gamma}^2/\hbar\alpha}\}.
\end{align}
\end{subequations}
Finally, if both the contributions are relevant we have
\begin{subequations}
\begin{align}
&\tilde{P}_1=1-\exp\{-2\pi\tilde{\Gamma}_+^2/\hbar\alpha\}, \\
&\tilde{P}_2=1-\exp\{-2\pi(\tilde{\gamma}^2+\tilde{\Gamma}_-^2)/\hbar\alpha\}.
\end{align}
\end{subequations}

We see that also the two-qutrit system presents different dynamical behaviours related to the presence or not of second order interaction terms like the DM and d-d couplings.
These different behaviours are visible in laboratory, that is, measurable dynamical effects which can be brought to light by applying an LMSZ ramp on the two spin-1's and by exploiting the coupling-based LMSZ transition effect.
The latter, therefore, reported and discussed in Refs. \cite{GVMqubits,GVMqutrits}, turns to be a useful instrument not only to generate both entangled state of the two qubits and the two qutrits.
This effect would in fact enable us to investigate physical aspects and characteristics of the system under scrutiny, like the type and the weight of the interactions existing between the two spins ultimately providing precious informations  to improve the same microscopic Hamiltonian model.

\subsection{Entanglement}\label{Sec IV-V}

In case of two qutrits the level of entanglement got established between the two subsystems can be studied through the concept of negativity introduced by G. Vidal and R. F. Werner in \cite{Vid-Wer}.
Mathematically it can be cast as follows \cite{Rai-Luthra}
\begin{equation}\label{General Negativity}
\mathcal{N}_\rho = {||\rho^{T_B} ||_1 - 1 \over 2},
\end{equation}
where $\rho$ is the two-qutrit density matrix, $\rho^{T_B}$ is its partial transpose with respect to the subsystem $B$ and $|| \cdot ||_1$ stands for the trace norm.
Therefore, for a hermitian matrix, the negativity turns out to be the sum of the absolute values of the negative eigenvalues.
As the concurrence for two qubits, $\mathcal{N}_{\rho}$ ranges from 0 to 1 \cite{Rai-Luthra}.
The choice of a necessarily factorized orthonormal basis does not affect the result as well as the subsystem with respect to which the partial transpose is calculated.

From Ref. \cite{GMIV} we know that the negativity in the four dimensional subspace is bounded from above and the limit value is $\mathcal{N}=1/2$.
In fact, for a generic pure state $\ket{\Psi}=c_1\ket{10}+c_2\ket{01}+c_3\ket{0-1}+c_4\ket{-10}$, the negativity simply reads \cite{GMIV}
\begin{equation}\label{Neg 4x4}
\mathcal{N}=\sqrt{x(1-x)}, \qquad x=|c_1|^2+|c_4|^2.
\end{equation}

In Ref. \cite{GVMqutrits}, instead, the asymptotic expression of the negativity is obtained in terms of the following asymptotic parameter
\begin{equation}\label{x}
x(\infty)=\tilde{P}_1\tilde{P}_2+(1-\tilde{P}_1)(1-\tilde{P}_2),
\end{equation} 
where $\tilde{P}_1$ and $\tilde{P}_2$ are the probability expressions of the two fictitious qubits reported in Eq. \eqref{Prob fict spin 1/2 qutrits}.
It is easy to verify \cite{GVMqubits} that the asymptotic negativity plotted in terms of the LMSZ parameter presents two maxima which correspond to the values $\log(2)/2\pi\approx 0.11$ and $\log(2)/\pi\approx 0.22$ of the LMSZ parameter, respectively.
It means that when the LMSZ parameter equals one of these two values the two qutrits asymptotically reach a state with the possible maximum level of entanglement allowed in the four dimensional subspace.

Therefore, analogously to the reasoning made before for the two-qubit system, it is easy to understand that for the two qutrits too the presence of DM and/or d-d interaction can result in an enhancement or reduction of the level of entanglement with respect to the dynamical situation wherein d-d and DM interactions are ignored.
It depends on the fact that the DM and/or d-d contribution can either help to reach one of the two magic values to generate entangled states or make the LMSZ parameter far from the same values.
We may say, thus, that this sensitivity can be a useful tool for measuring the DM coupling.
A similar analysis with analogous results can be developed for the three-dimensional subspace considered above in five-dimensional subspace under specific conditions on the relevant parameters.

%

\section{Conclusive Remarks}\label{Sec C}

In Refs. \cite{GVMqubits} and \cite{GVMqutrits} the authors brought to light a physical effect called coupling-based Landau-Majorana-St\"uckelberg-Zener (LMSZ) transitions for two-qubits and two-qutrit systems, respectively.
By applying an LMSZ ramp on the spin-system we can speak of joint LMSZ transition probabilities, even though a constant transverse field is absent.
This is possible thanks to the coupling existing between the spins.
An interesting aspect is that an exact analysis can be developed and analytical results can be obtained.
This circumstance relies on the fact that the symmetry properties possessed by the Hamiltonians allow us to identify dynamically invariant Hilbert subspaces and consequently to reduce the dynamical problem into relatively easier sub-dynamics.
In case of two qubits we end up with two two-dimensional subspaces and then we can solve the two-spin dynamics by solving separately the two two-level dynamical problems.
This dynamical decomposition approach was used to find other remarkable features of the two-qubit system \cite{GMGIM} and to study the exact dynamics of more complex systems like two qudits \cite{GBNM}, $N$-qubit chain \cite{GLSM} and pairs of interacting quantum harmonic oscillators \cite{GMMM}.
Moreover, it is important to underline that the dynamical decomposition method is independent of the specific time-dependent scenario we take into account.
Its more general validity, then, allows us to consider other exactly solvable scenarios \cite{Bagrov, KunaNaudts, Das Sarma, Mess-Nak, MGMN, GdCNM1, SNGM} leading us to new exactly solvable dynamics of the spin-systems.

On the basis of the previous results, in this work we examined possible experimentally detectable effects on the joint coupling-based LMSZ transitions stemming from the presence of Dzyaloshinskii-Moriya (DM) and/or dipole-dipole (d-d) interactions.
In Refs. \cite{GVMqubits} and \cite{GVMqutrits}, in fact, only the anisotropic exchange interaction was considered.
Here we demonstrate that the presence of DM and/or d-d interactions affects the joint coupling-based LMSZ transitions.
Since the interaction terms added to the models studied in Refs. \cite{GVMqubits} and \cite{GVMqutrits} do not break the symmetry properties of the Hamiltonians, we are able to treat exactly even the new dynamical problem.
Therefore also in this case we were able to disclose the spin-dynamics by solving lower dimensional dynamical problems.

We brought to light that, on the one hand, both interactions lead to an enhancement of the joint LMSZ transition probabilities, both for the spin-qubits and the spin-qutrits [see Eqs. \eqref{P+- gen} and \eqref{Prob fict spin 1/2 qutrits}, respectively].
On the other hand, DM (d-d) coupling produces physical effects different from those rising in the presence of d-d (DM) coupling.
For the two-qubit case, for example, when an isotropic exchange is considered, it governs the LMSZ dynamics in one of two dynamically invariant subspaces, while the d-d interaction makes the LMSZ transition possible in the other subspace [see Eq. \eqref{P iso change + dd}].
The DM interaction, instead, contributes to enhance the LMSZ transition probability ruled by the isotropic exchange interaction; in this case the LMSZ transition is hindered in the other sub-space [see Eq. \eqref{P iso change + DM}].
These results are important since, by studying the LMSZ transitions in the two subspaces (addressable by preparing the spin-system in the appropriate initial condition), we can get information about what kind of interaction characterizes the spin-system and what are the possible different relative weights of these interactions.

Another interesting aspect concerns the entanglement.
In Refs. \cite{GVMqubits} and \cite{GVMqutrits} the authors showed that the coupling-based LMSZ transitions can be exploited to generate entangled states of the two-qudit systems (precisely, maximally entangled states for the two qubits).
This is possible when a precise condition on the LMSZ parameter ruling the dynamics is fulfilled.
Here, we demonstrated that the presence of DM and/or d-d interaction can facilitate or impede the achievement of this condition for the entanglement.
Therefore, the two interactions considered in this work can generate either positive or negative effects on the entangled-state generation.
A curious aspect to point out is that when both DM and d-d are neglected, entanglement generation requires  non-adiabatic conditions, as already stressed in the previous works \cite{GVMqubits,GVMqutrits}.
In this paper we instead make evident that the entanglement generation can be realized under adiabatic and more `comfortable' conditions too.
This is possible thanks to the `half-crossing' scenario \cite{Carr-Hioe}, in which the linearly varying field is turned on exactly at the crossing point.
It ensures an asymptotic transition probability equal to 1/2 \cite{Vit-Garr,Vitpra99,Vit-Zlat}, which is exactly the key point to realize entangled states.
Neverthless, it turns out to be easier and more suitable to be implemented in experiments \cite{Yamazaki,Randall} than the full crossing LMSZ model under the specific necessary condition in Eq. \eqref{Condition max ent}.

Finally, a possible development of the present work could be the study of other detectable effects on the joint coupling-based LMSZ transitions.
It could be surely interesting to analyse how the interaction of the spin system with a surrounding environment affect the dynamics.
This kind of problem can be approached through at least three methods: the GKLS theory based on master equations \cite{GKLS}, the numerical approach based on the Wigner partial transpose \cite{Kapral,SHGM,SGHM} and the effective description in terms of non-Hermitian Hamiltonians \cite{Feshbach,Rotter,GdCKM,GdCNM}.

\section*{Acknowledgements}
NVV acknowledges support from the EU Horison-2020 project 820314 (MicroQC).
HN is partly supported by Waseda University Grant for Special Research projects 2019-C256.

\appendix

\section{Previous results}

\subsection{Qubits}

In Ref. \cite{GVMqubits} the authors consider the following two-qubit model
\begin{equation} \label{Hamiltonian old}
\begin{aligned}
{H} =
\hbar\omega_{1}(t)\hat{\sigma}_{1}^{z}+\hbar\omega_{2}(t)\hat{\sigma}_{2}^{z}+\gamma_{x}\hat{\sigma}_{1}^{x}\hat{\sigma}_{2}^{x}+\gamma_{y}\hat{\sigma}_{1}^{y}\hat{\sigma}_{2}^{y}+\gamma_{z}\hat{\sigma}_{1}^{z}\hat{\sigma}_{2}^{z}
\end{aligned}
\end{equation}
where $\hat{\sigma}_{i}^{x}$, $\hat{\sigma}_{i}^{y}$ and $\hat{\sigma}_{i}^{z}$ ($i=1,2$) are the Pauli matrices.
A symmetry-based analysis of the Hamiltonian model brings to the identification of two two-dimensional dynamically invariant Hilbert subspaces.
One is spanned by the two standard basis states $\{\ket{++},\ket{--}\}$ and its dynamics is governed by the following effective two-level Hamiltonian
\begin{equation}
H_+=\hbar\Omega_+(t)\hat{\sigma}^z+\gamma_+\hat{\sigma}^x+\gamma_z\mathbb{1}_+.
\end{equation}
The other subspace, instead, involves the two standard basis states $\{\ket{+-},\ket{-+}\}$ and the effective two-level Hamiltonian ruling its dynamics may be cast as
\begin{equation}
H_-=\hbar\Omega_-(t)\hat{\sigma}^z+\gamma_-\hat{\sigma}^x-\gamma_z\mathbb{1}_-.
\end{equation}
In the previous expressions $\mathbb{1}_\pm$ represent the identity operators within the two distinct subspaces and we put $\Omega_\pm=\omega_1\pm\omega_2$ and $\gamma_\pm=\gamma_x\mp\gamma_y$.
It is worth pointing out that, through this symmetry-based analysis, the two-qubit dynamical problem is reduced to the study and the solution of two independent two-level dynamical problems.

Thanks to the subdivision of the Hilbert space and the breaking-down of the dynamical problem, it is possible to construct the formal expression of the time evolution operator $U(t)$ related to the two-qubit Hamiltonian, that is solution of the Schr\"odinger equation $i\hbar\dot{U}(t)=H(t)U(t)$.
Depending on the time-dependence of the Hamiltonian parameter (in this case $\omega_1(t)$ and $\omega_2(t)$) it is possible to construct the specific exact expression if we are able to solve analytically the two single-qubit subdynamical problems. 

For example, by considering an LMSZ-like scenario, that is a magnetic field ramp applied on the first spin, namely $\omega_1(t)=\alpha~t$ and $\omega_2(t)=0$, we can exploit the LMSZ result to write down the two-qubit asymptotic transition probabilities
\begin{equation}\label{P+ P- simple}
\begin{aligned}
&P_+(\infty)=|\average{++|U_+(\infty)|--}|^2=1-\exp\{ -2\pi\gamma_+^2/\hbar\alpha \}, \\
&P_-(\infty)=|\average{+-|U_-(\infty)|-+}|^2=1-\exp\{ -2\pi\gamma_-^2/\hbar\alpha \},
\end{aligned}
\end{equation}
These expressions show that LMSZ-like transitions of the two-qubit system are possible although a transverse constant field is absent.
The role of the latter is indispensable for the occurrence of LMSZ transition in a single two-level system.
Nevertheless, such a role, in the case of the two interacting qubits, is played by the presence of the coupling between the spins and this is the reason why we can speak of coupling-based two-qubit LMSZ transitions.

It is important to point out that Eqs. \eqref{P+ P- simple} are exact but asymptotic expressions of the transition probabilities related to the LMSZ ideal model consisting in an infinite procedure.
However, if we are interested in considering a finite time window for the LMSZ procedure, we can exploit the exact solution of the dynamical problem reported in Ref. \cite{Vit-Garr}.
In this case we can write the analytical expressions and the exact time behaviour of the transition probabilities \cite{GVMqubits}.
This circumstance makes possible also the exact analysis of the time dependence of the entanglement got established between the two qubits \cite{GVMqubits}.
Moreover, in Ref. \cite{GVMqubits} the authors brought to light a possible application based on the coupling-assisted LMSZ transitions.
They showed, indeed, the possibility of generating maximally entangled states of the two spin-qubits by appropriately setting the ratio between the field's slope and the coupling parameter.

\subsection{Qutrits}

In Ref. \cite{GVMqutrits} the authors considered the same model as in Eq. \eqref{Hamiltonian old} for two qutrit-spins.
The symmetries possessed by the Hamiltonian model are independent of the value of the two interacting spins.
This fact implies that also for two qutrits we can identify two dynamically invariant Hilbert subspaces and describe the two-qutrit dynamics within each subspace in terms of fictitious systems.
Precisely, in the two-qutrit case, the Hilbert space is decomposed in a four-dimensional and a five-dimensional subspaces.
The former is spanned by $\{ \ket{10}, \ket{01}, \ket{-10}, \ket{0-1} \}$ and is characterized by a two-qutrit dynamics which can be effectively described in terms of two decoupled fictitious spin-1/2's subjected to the following two single-qubit Hamiltonians
\begin{equation}\label{H1 and H2}
H_{1}=\frac{\hbar\Omega_+}{2}\hat{\sigma}_{1}^{z}+
\tilde{\gamma}_-\hat{\sigma}_{1}^{x},
\qquad
H_{2}=\frac{\hbar\Omega_-}{2}\hat{\sigma}_{2}^{z}+
\tilde{\gamma}_+\hat{\sigma}_{2}^{x}
\end{equation}
with $\tilde{\gamma}_\pm=\gamma_x\pm\gamma_y$.
This circumstance allows to construct easily the time evolution operator ruling the dynamics in such a subspace and to study the exact two-qutrit behaviour under the LMSZ scenario.
We can speak of coupling based LMSZ this time too and we have precisely
\begin{equation}\label{MLSZ Trans Prob 0-1 to 01}
\begin{aligned}
|\average{10|U_+(\infty)|-10}|^2 &= P_1 P_2, \\
|\average{01|U_+(\infty)|-10}|^2 &= P_1 (1-P_2), \\
|\average{0-1|U_+(\infty)|-10}|^2 &= (1-P_1) P_2,
\end{aligned}
\end{equation}
with
\begin{equation}\label{P1}
P_1=1-\exp\{-2\pi\tilde{\gamma}_-^2/\hbar\alpha\}, \qquad P_2=1-\exp\{-2\pi\tilde{\gamma}_+^2/\hbar\alpha\}.
\end{equation}

The second subspace, under specific conditions on the coupling parameters (namely $\gamma_x=\gamma_y=\tilde{\gamma}/2$) can be reduced into two unidimensional Hilbert subspaces and a three-dimensional one.
The latter is spanned by the states $\{ \ket{1-1}, \ket{00}, \ket{-11} \}$ and can be described in terms of a single fictitious qutrit subjected to the following Hamiltonian
\begin{equation}\label{H3}
H_3=
\tilde{\gamma}~\hat{\Sigma}^{x}+\hbar\Omega_-\hat{\Sigma}^{z}.
\end{equation}
Also in this case, thus, it is possible to obtain the exact expressions for the LMSZ transitions, namely
\begin{equation}\label{LMSZ Tr Pr 3-lev}
\begin{aligned}
P_{-1}^{+1}=P_3^2, \quad
P_{-1}^{0}=2P_3(1-P_3), \quad
P_{-1}^{-1}=(1-P_3)^2,
\end{aligned}
\end{equation}
where $P_3=1-\exp\{-4\pi\tilde{\gamma}^2/\hbar\alpha\}$.
Finally, by studying the exact time behaviour of the negativity, the authors have shown that, exploiting the coupling-based LMSZ transitions, it is possible to generate entangled states of the two qutrits too.


\begin{thebibliography}{99}

\bibitem{Dzyaloshinskii}
I. Dzyaloshinskii, J. Phys Chem. Solids \textbf{4}, 241 (1958).

\bibitem{Moriya}
T. Moriya, Phys. Rev. \textbf{120}, 91 (1960).

\bibitem{Dender}
D. C. Dender, P. R. Hammar, D. H. Reich, C. Broholm, and G. Aeppli, Phys. Rev. Lett. \textbf{79}, 1750 (1997).

\bibitem{Kohgi}
M. Kohgi, K. Iwasa, J. M. Mignot, B. Fak, P. Gegenwart, M. Lang, A. Ochiai, H. Aoki, and T. Suzuki, Phys. Rev. Lett.
\textbf{86}, 2439 (2001).

\bibitem{Tsukada}
I. Tsukada, J. T. Takeya, T. Masuda, and K. Uchinokura, Phys. Rev. Lett. \textbf{87}, 127203 (2001).

\bibitem{Caretta}
S. Carretta et al., Phys. Rev. Lett. \textbf{100}, 157203 (2008).

\bibitem{Ramsey}
C. M. Ramsey et al., Nature Phys. \textbf{4}, 277 (2008).

\bibitem{Kavokin}
K. V. Kavokin, Phys. Rev. B \textbf{64}, 075305 (2001).

\bibitem{Wu}
L.-A. Wu and D. A. Lidar, Phys. Rev. A \textbf{66}, 062314 (2002); L.-A. Wu and D. A. Lidar, Phys. Rev. Lett. \textbf{91}, 097904 (2003).

\bibitem{Jafari}
R. Jafari,M. Kargarian,A. Langari, and M. Siahatgar, Phys. Rev. B \textbf{78}, 214414 (2008); M. Kargarian, R. Jafari, and A. Langari, Phys. Rev. A \textbf{79}, 042319 (2009).

\bibitem{Wang}
B. Wang, M. Feng, and Z.-Q. Chen, Phys. Rev. A \textbf{81}, 064301 (2010).

\bibitem{Kwan}
M. K. Kwan, Zeynep Nilhan Gurkan, and L. C. Kwek, Phys. Rev. A \textbf{77}, 062311 (2008).

\bibitem{Wernsdorfer}
W. Wernsdorfer, T. C. Stamatatos, and G. Christou, Phys. Rev. Lett. \textbf{101}, 237204 (2008).

\bibitem{LiuKong}
Fu-Wu Ma, Sheng-Xin Liu, and Xiang-Mu Kong, Phys. Rev. A \textbf{84}, 042302 (2011).

\bibitem{Maruyama}
Koji Maruyama, Toshiaki Iitaka, and Franco Nori, Phys. Rev A \textbf{75}, 012325 (2007).

\bibitem{Zhang}
Guo-Feng Zhang, Phys. Rev. A \textbf{75}, 034304 (2007).

\bibitem{Liu}
Ben-Qiong Liu, Bin Shao, Jun-Gang Li, Jian Zou, and Lian-Ao Wu, Phys. Rev. A \textbf{83}, 052112 (2011).

\bibitem{Sahling}
S. Sahling, G. Remenyi, C. Paulsen, P. Monceau, V. Saligrama, C. Marin, A. Revcolevschi, L. P. Regnault, S. Raymond and J. E. Lorenzo, Nat. Phys. \textbf{11}, 255-260 (2015).

\bibitem{Jurcevic}
P. Jurcevic, B. P. Lanyon, P. Hauke, C. Hempel, P. Zoller, R. Blatt and C. F. Roos, Nature \textbf{511}, 202-205 (2014).

\bibitem{Richerme}
P. Richerme, Zhe-Xuan Gong, A. Lee, C. Senko, J. Smith, M. Foss-Feig, S. Michalakis, A. V. Gorshkov and C. Monroe, Nature \textbf{511}, 198-201 (2014).

\bibitem{Boness}
T. Boness, S. Bose, T. S. Monteiro, Phys. Rev. Lett. \textbf{96}, 187201 (2006).

\bibitem{Amico}
L. Amico, R. Fazio, A. Osterloh, and V. Vedral, Rev. Mod. Phys. \textbf{80}, 517 (2008).

\bibitem{Arnesen}
M. C. Arnesen, S. Bose, V. Vedral, Phys. Rev. Lett. \textbf{87}, 017901 (2001).

\bibitem{GVMqubits}
R. Grimaudo, N. V. Vitanov, and A. Messina Phys. Rev. B \textbf{99}, 174416 (2019).

\bibitem{LMSZ}
L. D. Landau, Phys. Z. Sowjetunion \textbf{2}, 46 (1932);
E. Majorana, Nuovo Cimento \textbf{9}, 43 (1932);
E. C. G. St\"uckelberg, Helv. Phys. Acta \textbf{5}, 369 (1932);
C. Zener, Proc. R. Soc. London, Ser. A \textbf{137}, 696 (1932).

\bibitem{Shevchenko1}
S.N. Shevchenko, S. Ashhab, Franco Nori, Phys. Rep. \textbf{492}, 1 (2010).

\bibitem{Vit-Garr}
N. V. Vitanov and B. M. Garraway, Phys. Rev. A \textbf{53}, 6 (1996).

\bibitem{Akulin}
V. M. Akulin and W. P. Schleich, Phys. Rev. A \textbf{46}, 7 (1992).

\bibitem{Vitanov}
N. V. Vitanov and S. Stenholm, Phys. Rev. A \textbf{55}, 4 (1997).

\bibitem{Pok-Sin2003}
V. L. Pokrovsky and N. A. Sinitsyn, Phys Rev. B \textbf{67}, 144303 (2003).

\bibitem{Ivanov}
P. A. Ivanov and N. V. Vitanov, Phys. Rev. A \textbf{71}, 063407 (2005).

\bibitem{Militello2}
M. Scala, B. Militello, A. Messina and N. V. Vitanov, Phys. Rev. A \textbf{84}, 023416 (2011).

\bibitem{Militello3}
A. V. Dodonov, B. Militello, A. Napoli, and A. Messina, Phys. Rev. A \textbf{93}, 052505 (2016).

\bibitem{Petta}
J. R. Petta  \textit{et al.}, Science \textbf{309} (5744) 2180-2184 (2005).

\bibitem{Anderlini}
M. Anderlini, P. J. Lee, B. L. Brown, J. Sebby-Strabley, W. D. Phillips, and J. V. Porto Nature \textbf{448} (7152) 452-456 (2007).

\bibitem{Foletti}
H. Bluhm, S. Foletti, I. Neder, M. Rudner, D. Mahalu, V. Umansky and A. Yacoby Nat. Phys. \textbf{7} 109 (2011).

\bibitem{Das Sarma Nat}
Xin Wang, L. S. Bishop, J. P. Kestner, E. Barnes, Kai Sun and S. Das Sarma Nat. Comm. \textbf{3} 997 (2012).

\bibitem{GMIV}
R. Grimaudo, A. Messina, P. A. Ivanov and N. V. Vitanov J. Phys. A \textbf{50} 175301 (2017).

\bibitem{GVMqutrits}
R. Grimaudo, N. V. Vitanov, and A. Messina Phys. Rev. B \textbf{99}, 214406 (2019).

\bibitem{Klimov2003}
A. B. Klimov, R. Guzman, J. C. Retamal, and C. Saavedra, Phys. Rev. A \textbf{67}, 062313 (2003).

\bibitem{Vitanov2012}
N. V. Vitanov, Phys. Rev. A \textbf{85}, 032331 (2012).

\bibitem{Ivanov2006}
P. A. Ivanov, E. S. Kyoseva, and N. V. Vitanov, Phys. Rev. A \textbf{74}, 022323 (2006).

\bibitem{Cerf2002}
N. J. Cerf, M. Bourennane, A. Karlsson, and N. Gisin, Phys. Rev. Lett. \textbf{88}, 127902 (2002).

\bibitem{Vaziri2002}
A. Vaziri, G. Weihs, and A. Zeilinger, Phys. Rev. Lett. \textbf{89}, 240401 (2002).

\bibitem{Bolton}
J. A.Weil and J. R. Bolton, Electron Paramagnetic Resonance - Elementary Theory and Practical Applications, 2nd ed. (Wiley, Hoboken, NJ, 2007).

\bibitem{GMN}
R. Grimaudo, A. Messina, H. Nakazato, Phys. Rev. A \textbf{94}, 022108 (2016).

\bibitem{Wootters}
W. K. Wootters, Phys. Rev. Lett. \textbf{80}, 2245 (1998).

\bibitem{Carr-Hioe}
C. E. Carroll and F. T. Hioe, J. Phys. A \textbf{19}, 1151 (1986).

\bibitem{Vitpra99}
N. V. Vitanov, Phys. Rev. A \textbf{59}, 988 (1999).

\bibitem{Vit-Zlat}
K. N. Zlatanov, N. V. Vitanov, Phys. Rev. A \textbf{96}, 013415 (2017).

\bibitem{Yamazaki}
R. Yamazaki, K. Kanda, F. Inoue, K. Toyoda, and S. Urabe, Phys. Rev. A \textbf{78}, 023808 (2008).

\bibitem{Randall}
J. Randall, A. M. Lawrence, S. C. Webster, S. Weidt, N. V. Vitanov, and W. K. Hensinger, Phys. Rev. A \textbf{98}, 043414 (2018).

\bibitem{Vid-Wer}
G. Vidal, R. F. Werner, Phys. Rev. A \textbf{65}, 032314 (2002).

\bibitem{Rai-Luthra}
S. Rai and J. R. Luthra, arXiv:quant-ph/0507263.

\bibitem{GMGIM}
R. Grimaudo, T. Mihaescu, I. Ghiu, A. Isar, A. Messina, Results in Physics {\bf 13}, 102147 (2019)

\bibitem{GBNM}
R. Grimaudo,Y. Belousov, H. Nakazato and A. Messina, Ann. Phys. (NY) \textbf{392}, 242 (2017).
    
\bibitem{GLSM}
R. Grimaudo, L. Lamata, E. Solano, A. Messina, Phys. Rev. A \textbf{98}, 042330 (2018).

\bibitem{GMMM}
R. Grimaudo, V. I. Man'ko, M. A. Man'ko, and A. Messina, Phys. Scr. \textbf{95} (2), 024004 (2019).

\bibitem{Bagrov}
V. G. Bagrov, D. M. Gitman, M. C. Baldiotti and A. D. Levin, Ann. Phys. (Berlin) \textbf{14} (11) 764 (2005).

\bibitem{KunaNaudts}
M. Kuna and J. Naudts Rep. Math. Phys. \textbf{65} (1) 77 (2010).

\bibitem{Das Sarma}
E. Barnes and S. Das Sarma Phys. Rev. Lett. \textbf{109} 060401 (2012).

\bibitem{Mess-Nak}
A. Messina and H. Nakazato J. Phys. A: Math. Theor. \textbf{47} 445302 (2014).

\bibitem{MGMN}
L. A. Markovich, R. Grimaudo, A. Messina and H. Nakazato Ann. Phys. (NY) \textbf{385} 522 (2017).

\bibitem{GdCNM1}
R. Grimaudo, A. S. M. de Castro, H. Nakazato and A. Messina, Ann. Phys. (Berlin) \textbf{530}, 12 1800198 (2018).

\bibitem{SNGM}
T. Suzuki, H. Nakazato, Roberto Grimaudo and Antonino Messina, Sc. Rep. \textbf{8}, 17433 (2018).

\bibitem{GKLS}
V. Gorini, A. Kossakowski, and E. C. G. Sudarshan, J. Math. Phys. \textbf{17}, 821 (1976);
G. Lindblad, Commun. Math. Phys. \textbf{48}, 119 (1976).

\bibitem{Kapral}
R. Kapral, G. Ciccotti, J. Chem. Phys. \textbf{110}, 8919-8929 (1999);
R. Kapral, J. Phys. Chem. A \textbf{105}, 2885-2889 (2001);
S. Nielsen; R. Kapral, G. Ciccotti, J. Chem. Phys. \textbf{115}, 5805-5815 (2001).

\bibitem{SHGM}
A. Sergi, G. Hanna, R. Grimaudo and A. Messina, Symmetry, \textbf{10}(10), 518 (2018).

\bibitem{SGHM}
A. Sergi, R. Grimaudo, G. Hanna, and A. Messina, Physics 1 (3), 402-411 (2019).

\bibitem{Feshbach}
H. Feshbach, Ann. Phys. \textbf{5}, 357 (1958); H. Feshbach, Ann. Phys. \textbf{19}, 287 (1962).

\bibitem{Rotter}
I. Rotter and J. P. Bird, Rep. Prog. Phys. \textbf{78}, 114001, (2015);
I. Rotter, J. Phys. A: Math. Theor. \textbf{42}, 153001, (2009).

\bibitem{GdCKM}
R. Grimaudo, A. S. M. de Castro, M. Ku\'s, and A. Messina, Phys. Rev A \textbf{98}, 033835 (2018).

\bibitem{GdCNM}
R. Grimaudo, A. S. M. de Castro, H. Nakazato, and A. Messina, Phys. Rev. A \textbf{99} (5), 052103 (2019).


\end{thebibliography}
\end{document}